\documentclass[aps,prl,twocolumn,10 pt,superscriptaddress,showpacs]{revtex4}
\usepackage{amssymb}
\usepackage{graphicx}

\begin{document}

\title{Magnetic states and spin-glass properties in $Bi_{0.67}Ca_{0.33}MnO_{3}$: macroscopic $ac$ measurements and neutron scattering.}
\author{Maud Giot}
\altaffiliation[Also at ]{SPMS, UMR CNRS 8580, École Centrale Paris, 92295 Châtenay-Malabry, France
 and DEN/DMN/SRMA/LA2M, CEA Saclay, 91191 Gif-sur-Yvette, France.}
\affiliation{Laboratoire CRISMAT, UMR 6508 du CNRS, ENSICAEN et Universit\'{e} de Caen, 6 Bd Mar\'{e}chal Juin, F-14050 Caen 4.}
\affiliation{Laboratoire L\'{e}on Brillouin,  CEN Saclay, 91191
Gif/Yvette, France.}
\author{Alain Pautrat}
\affiliation{Laboratoire CRISMAT, UMR 6508 du CNRS, ENSICAEN et Universit\'{e} de Caen, 6 Bd Mar\'{e}chal Juin, F-14050 Caen 4.}
\author{Gilles Andr\'{e}}
\affiliation{Laboratoire L\'{e}on Brillouin,  CEN Saclay, 91191
Gif/Yvette, France.}
\author{Damien Saurel}
\affiliation{Nanoscale Superconductivity and Magnetism and Pulsed Fields Group, Institute for Nanoscale Physics and Chemistry (INPAC), K. U. Leuven, Celestijnenlaan 200 D, 3001 Leuven, Belgium.}
\author{Maryvonne Hervieu}
\affiliation{Laboratoire CRISMAT, UMR 6508 du CNRS, ENSICAEN et Universit\'{e} de Caen, 6 Bd Mar\'{e}chal Juin, F-14050 Caen 4.}
\author{Juan Rodriguez-Carvajal}
\affiliation{Institut Laue-Langevin, 6, rue Jules Horowitz, 
BP 156, 38042 Grenoble Cedex 9, France}

\begin{abstract}
We report on the magnetic properties of the manganite $Bi_{1-x}Ca_{x}MnO_3$ ($x=0.33$) at low temperature. The analysis of
the field expansion of the $ac$ susceptibility and the observation of aging properties make clear that a
spin glass phase appears below $T = 39K$, in the presence of magnetic order. Neutron scattering shows both magnetic Bragg scattering and
 magnetic diffusion at small angles, and confirms this coexistence.
In contrast to $Pr_{1-x}Ca_{x}MnO_3$ ($x=0.3-0.33$) which exhibits a mesoscopic phase separation responsible for a field driven percolation, the glassy and short range ferromagnetic order observed here does not cause colossal magnetoresistance (CMR).

\end{abstract}

\newpage
\maketitle

I/INTRODUCTION\\

Strong efforts have been made recent years to improve the
understanding of the properties of manganese oxides. It has
been clarified that a strong tendency exists to form mixed states, when different degrees of freedom, magnetic interactions and disorder effects are in competition \cite{dagotto}. An experimental
 consequence is that magnetic inhomogeneities coexisting
over several scales can be observed.
 A particular interest can be found when
working with compositions which are just between dominant ferromagnetic and
antiferromagnetic interactions. Indeed, when the contribution of
ferromagnetic and antiferromagnetic exchanges compare, magnetic
long range ordering can be frustrated and magnetic phases show peculiar properties: A spin glass
is one of the most celebrated examples. Some frustrated
magnets exhibit the coexistence of order in some sub-lattices with disorder in other sub-lattices \cite{order}.

Among different manganites, the Bi-rich part of the phase diagram of $Bi_{1-x}Ca_{x}MnO_{3}$ has been scarcely investigated. It was done some years ago by \textit{Bokov et al}, and more recently by
\textit{Troyanchuk et al} \cite{bokov,troy}. At low temperature, for $x
\leq  0.15$, ferromagnetic ordering occurs and for $x \geq  0.35$, the ground state is antiferromagnetic. Between these values, the
situation can be complex due to the competition between magnetic interactions. In addition, the extreme sensitivity to
the cationic or oxygen contents plays a role for properties, with the possible existence of critical compositions.
 For $x = 0.32$, using field
cooling (FC) and zero field cooling (ZFC) magnetic measurements, a
spin-glass phase has been proposed for $T \lesssim 40 K$ \cite{troy}.
 Nevertheless, there is $\textit{a priori}$ no need for spins freezing to have hysteretic properties. 
Thermal hysteresis is observed in canonical spin-glasses, but also in conventional ferromagnets
  which can exhibits a broad distribution of potential barriers, in superparamagnets (...). Moreover, canted ferromagnetic states cannot be excluded \cite{degennes}. The low temperature properties of this phase remains to be investigated in details.

 To say more on the nature of the magnetic states, the non-linear susceptibility
$\chi_{nl}$ is an interesting tool. In particular, the spin-glass transition can be robustly characterized
by a divergence of the cubic component of the magnetization, which occurs at the temperature of spin-glass order \cite{binder}.
$\chi_{nl}$ measurements has been less used for proving ferromagnetic
or antiferromagnetic transitions \cite{nail}, but allows to evidence an incipient magnetic order (Note that this technique is more usually employed
 for proving a paraelectic to ferroelectric transition, when analyzing the high order harmonics of the electric susceptibility).
A particular interest can be found in the case of multicomponent
magnetism where, with the linear susceptibility only, it is often
difficult to prove a phase coexistence. Furthermore, since $dc$ measurements do not probe kinetic effects which are essential to get the proof of glassy properties,
an analysis of $ac$ magnetic properties is complementary. In the case of manganites, it is particularly important to focus on the nature of the disordered state. For an equivalent doping, $Pr_{1-x}Ca_xMnO_3$ (x=0.30-0.33) shows intriguing properties. 
The competition between ferromagnetic and antiferromagnetic interactions
leads to a phase separation at a quasi-mesoscopic scale. This phase separation seems to be responsible 
for the properties of colossal magneto-resistance (CMR), via a mechanism of percolation \cite{damien}. In our sample, we have measured no CMR effect, at least up to 7 Teslas,
 and it is thus interesting to have some information on the nature of the disorder.
 
In this paper, we present a detailed study of the magnetic properties of $Bi_{0.67}Ca_{0.33}MnO_{3}$ at low temperatures. This compound is at the frontier between long range ferromagnetic and anti-ferromagnetic regimes, so that a complex coexistence takes place at the sample scale. The linear and non linear ac responses show the coexistence of
of magnetic behaviors take place at different scales in the sample. Neutrons scattering data evidence magnetic Bragg scattering coexisting with small angle diffusion, whose temperature dependence indicates a magnetic origin.
 This confirms the simultaneous existence of order and disorder states of the spins.
\\

II/EXPERIMENTAL\\

The polycrystalline sample $ Bi_{0.67}Ca_{0.33}MnO_{3} $ was prepared by solid 
state reaction. The appropriate proportion of $Bi_{2}O_{3}$, $CaO$, $Mn$ and $MnO_{2}$ powders were intimately mixed in an agate mortar. The mixture was then pressed in the form of parallelepipeds (2.5 $\times$ 2 $\times$ 10 
$mm^{3}$) under 1 $t/cm^{2}$. The parallelepipeds were disposed in an aluminum 
oxide finger and sealed in silica tubes under vacuum. To improve the crystalline 
quality of the ceramic, an oxygen pressure was imposed in the sealed tube by 
calculating the sample composition with 3.2 oxygen per formula. At 1050 $^{\circ} C$, this 
pressure is about 0.5 $MPa$. The samples were heated at 1050 $^{\circ} C$ for 24 hours then cooled 
for 6 hours. The quality of each sample was systematically checked. This was made by X-ray diffraction using a 
PHILIPS 1830 diffractometer with Cu K$\alpha$ radiation. Crystalline quality was also
checked by electron diffraction and cation composition homogeneity by EDS on 
several cristallites ($\cong$ 20) for each sample. 
The composition deduced from the EDS measurement is $Bi_{0.67(3)}Ca_{0.33(3)}MnO_{3}$.
 The uncertainty roughly reflects the intrumental resolution, showing a good homogeneity of the samples.
Nevertheless, it should be emphasized that the statistical distribution of cations
is unavoidable. It becomes really important to account for it when dealing with composition close to the frontier between two regimes in a phase diagram, as it is the case here.                                                 
One of the effect is that cationic inhomogeneities can coexist at small scales:
both charge ordered (CO) and non CO states, associated respectively to Bi-poor and Bi-rich phases,
have been seen in a crystallite of about $100$ nm by electron diffraction and transmission electron microscopy.

Seven sealed tubes, each containing one gram of the mixture, were 
prepared to obtain enough powder for neutron diffraction.  Neutron Powder Diffraction (NPD) data, for the magnetic scattering studies, were collected at Laboratoire L\'{e}on Brillouin on the G4.1 diffractometer ($\lambda$ = 2.425 \AA ,  2.00$^{\circ} \leq 2\theta  \leq$ 82.00$^{\circ}$ with a step of 0.10 degree) in the range of temperature 1.5 $\leq T \leq$ 300 $K$ (step 10 $K$ between 1.5 $K$ and 300 $K$).                                           
The ac and dc susceptibility measurements were performed using a
commercial superconducting-quantum-interference-device (SQUID, 
MPMS quantum design). The applied $ac$ field was 1 $Oe$ in order to be in the linear $ac$ regime (this was carefully checked), with a frequency ranging from 0.7 $Hz$ to 1 $KHz$. Low to moderate $dc$ magnetic fields (0-1000 $G$) were applied.\\

III/ MAGNETIZATION MEASUREMENTS\\

The $dc$ susceptibility exhibits Curie-Weiss behavior $\chi=C/(T-\theta_P)$ at high
temperature, with dominant ferromagnetic fluctuations characterized by a
positive $\theta_P \approx 20 K$ (fig.1), as previously reported \cite{troy,maud}. 
A deviation from the Curie-Weiss law when the temperature decreases can be observed for $T \leq 140 K$. Such a rounding of the susceptibility curve is a typical feature of magnetic
clustering \cite{mydosh}. More precisely, growing antiferromagnetism will be proved by neutron scattering, and will be discussed below. A thermal hysteresis between field cooled and zero field cooled
magnetization appears when $T_{max} \lesssim 39K $.

At low temperature ($T = 2 K$, fig. 2), the
magnetization versus applied magnetic field has a S-shape. A small hysteresis
can also be observed, with a crossing of the up and down branches. This latter
feature is typical of slow
relaxing states (superparamagnetism, disordered ferromagnetism
with some broad distribution of barriers, spin glasses.. ). The temperature dependence of
the field cooling magnetization is not in
agreement with superparamagnetism. This latter is more characterised by an increase of magnetization when the temperature
decreases, because of the thermal origin of the blocking of super
spins \cite{saza}. In contrast, we observe here a relative flat
field-cooled magnetization below a temperature $T_{max} \approx  39 K$.

Disordered systems can be differentiated if one studies their dynamic properties. If the kinetic is slow, 
low frequency $ac$ measurements
are well adapted. Primary characteristics are depicted in
the frequency dependence of the maximum of $\chi^{'}(T)$. Here, we observe the following
characteristics:
 (i) A very small frequency dependence of the temperature of the cusp.
In addition, a small depression of the peak of the order of $1 \%$ is observed if the frequency $f$ is changed from $1 Hz$ to $1 KHz$ ac field.
A phenomenological classification can be obtained through the quantity $\Delta T / T_{max} \Delta (ln 2 \pi f)$.
 For a conventional superparamagnet, a high
sensitivity to the frequency should be observed, evidencing blocking of
independent magnetic clusters. For spin-glasses, cooperativity
weakens this dependence. One finds 0.003 in the present case.
(ii) the onset of dissipation proved by a non zero value of $\chi^{"}$ near the
same temperatures.
 (iii) the collapse of the dissipation peak for moderate
applied magnetic field, evidencing strong sensitivity to a magnetic polarisation (see fig 3). 

(i), (ii) and (iii) are characteristics of a
spin-glass transition \cite{binder}.
We have also observed a logarithmic frequency
dependence in the susceptibility spectrum $\chi (f)$ for $T
< T_{max}$ (see fig.4). Such dependence can be found in the linear
response of glassy states, using a
classical two-level model with a broad distribution of random
potential \cite{natterman,pytte}.
They are thus several clues to associate $T_{max}$ with a spin-glass transition.

A more robust proof can arise from the behavior of the non-linear
susceptibility $\chi_{nl}$ close to the transition temperature
\cite{binder}. Theoretically, $\chi_{nl}$ was
 shown to mimic the behavior of the
 Edward-Anderson (order) parameter of spin glasses, and a negative divergence
	is expected in the zero field limit \cite{suzuki}.
 In practice, a rather sharp peak is
taken as a good proof of the spin-glass transition. This is to be compared with the broad peak observed in the case of superparamagnets \cite{peak}.

To analyze the non-linear susceptibility, we have measured the low
field expansion of the $ac$ susceptibility. Generally, the
magnetization can be expanded in power of the magnetic
field H (for small fields):

\begin{equation}
M = \chi_0.H + \chi_1.H^2 + \chi_2.H^3 + \chi_3.H^4+.....
\end{equation}

In our $ac$ measurements, a small field $h_0 exp(i \omega t)$ of typically $1 oe$ is
applied. In the low field limit $h_0 << H$, one can approximate respectively the ac
magnetization $M (\omega)$ and the ac susceptibility $\chi
(\omega) = \frac{\partial M(\omega)}{\partial h}$ by \cite{levy}:

\begin{equation}
M (\omega) = \chi_0.h_0 + 2 \chi_1.H.h_0 + 3 \chi_2.H^2.h_0 + 4
\chi_3.H^3.h_0+.....
\end{equation}

\begin{equation}
\chi (\omega) =  \chi_0 + 2 \chi_1.H + 3 \chi_2.H^2 + 4
\chi_3.H^3+.....
\end{equation}

For paramagnetic or spin-glass states, no spontaneous moment
are present, $\chi$ is unchanged by the symmetry $H \rightarrow -H$
and only odd terms can exist in equation (1).

\begin{equation}
\chi (\omega^{'}) =  \chi_0 + 3 \chi_2.H^2 + 5 \chi_4 H^4 +.....
\end{equation}

The high order terms of the expansion are usually negligible far
from a transition but become important close to the critical regime.
 To
analyze the non linear part $\chi^{'}_{nl} = \chi (\omega^{'}) -
\chi_0$, we have measured the $ac$ susceptibility as function of
the magnetic field (from 0 to 500 G) near $T_{max}$.
 As shown in the figure 5, when the temperature is slightly higher than $T_{max}$, the
quadratic magnetic field dependence of $\chi (\omega^{'})$ is
obvious. $\chi (\omega^{'})$ is symmetric when the magnetic
field is reversed, in agreement with the equation (4). A fitting of the data
allows to extract $\chi_2$ whose temperature variation is shown in
the fig. 6. Clearly, a peak of $\chi_2$ is observed at
$T_{max}$. If $-\chi_2$ is plotted as function of $\epsilon = T-T_{max}/T_{max}$ in a log-log scale, a linear part is observed for $\epsilon \geq 0.1$ (fig.7).
 It is consistent with a critical relation $\chi_2=\epsilon^{-\gamma }$, with here $\gamma= 2.4 \pm 0.1$. A similar exponent has been 
 observed in canonical spin-glasses like $Ag(Mn)$ \cite{levy}.
 We note also that the typical $T^{-3}$ dependence of $-\chi_2$, predicted in the Wohlfarth blocking model of superparamagnets, can not describe our data \cite{peak}.
Finally, from this analysis, it can be concluded that a spin glass transition occurs at $T_{sg}=T_{max}$.

Nevertheless, a peculiar behavior can
be observed for $T < T_{sg}$. The magnetic field dependence of
$\chi^{'} (\omega)$ becomes clearly non symmetric if one reverses the magnetic field (fig.5). This implies that the odd terms in equation (1)
($\chi_1, \chi_3, ...$) should be taken into account. 
 We emphasize that these terms are zero in a purely disordered state. 
Here, the symmetry breaking by field inversion
implies the existence of magnetic order in the spin glass state (Fig.6).
One has to note that our procedure which uses the field expansion does not allow
 to analyze the true zero field non-linear susceptibility. 
Thus, one can not completely exclude that this magnetic order comes from the incipient polarisation of magnetic clusters,
i.e. is not present in the zero field limit.
Anyway, a broad peak appears in the imaginary part of the ac susceptibility at the same temperature (fig.3), that can be naturally attributed to an extra-dissipation in the ordered state (magnetic domain walls).
Since this peak is almost insensitive to the magnetic field, we think that this is a robust indication that the ordered magnetic component exists also in the zero field limit. 

A known case of spin glass in competition with magnetic order is the reentrant spin glass.
In this case, the temperature of long range ordering is higher than the temperature of spin glass transition and the disorder state appears at the place of the ordered one (two distinct peaks in the temperature dependence of the coefficient of the
 $\chi_2$ term in the susceptibility can be observed \cite{reent}). This is quite different from our situation where magnetic order and disorder seems to coexist in the same temperature range.

One can wonder if typical spin glass properties can be observed even in presence of this magnetic order.
We have thus performed time-dependent $ac$ susceptibility in the linear regime.
In spin-glasses, after a temperature quench down to the low temperature state,
the susceptibility relaxes very slowly in a manner that depends on
the time $t_{\omega}$ spent before cutting the field. In particular, the
relaxation becomes slower as $t_{\omega}$ is longer, showing that aging
occurs \cite{vincent}. If $dc$ measurements are performed, a rescaling in $(t-t_{\omega})^{\beta }$ can be observed. In the case of ac measurements, 
the equivalent of the waiting time is now the measuring time, which is the inverse of the frequency $f^{-1}=(\omega / 2\pi)^{-1}$.
 The scaling becomes actually more simple, in a form $\omega t$. This is what we have observed here for $T = 34 K$ and for $T = 10 K$ (fig.8).
 Using temperature-cycling experiments (not shown here), the aging is observed as cumulative, indicating that our system behaves like a so-called superspin glass  \cite{reju}.
This shows that, in this sample, the glassy properties are dominating the macroscopic magnetic properties even in the presence of long range magnetic order \cite{ladieu}. 

Finally, using macroscopic measurements, magnetic order and disorder are found to coexist
at low temperature. Some examples of similar behavior are quantum-Heisenberg antiferromagnets \cite{quantum} and dilute Ising antiferromagnets \cite{wong}.\\

IV/ MAGNETIC NEUTRON SCATTERING: EVIDENCE OF LONG RANGE ORDER COEXISTING WITH DISORDER.\\

\indent A direct confirmation of magnetic order coexisting with magnetic disorder in \mbox{Bi$_{0.67}$Ca$_{0.33}$MnO$_{3}$} is provided by
neutron scattering. A diffractogram collected at 1.5K is shown in the fig.9. and is compared with the Rietveld refinement.
 Note that a strong small angle intensity ($\theta  \leq 10.5^{\circ}$) is also observed.
This latter can be not considered in the Rietveld analysis and its significance will be discussed afterwards.
To model the complex magnetic state, the Rietveld refinement has been performed considering two phases: 
\begin{enumerate}
	\item a charge ordered phase (orthorhombic cell, $a_{P}\sqrt{2}*2a_{P}\sqrt{2}*2a_{P}$ with 
	$a_{P}\simeq 3.9$\AA\ the cubic perovskite cell parameter, space group \textit{P2$_1$nm}) 
	associated to an antiferromagnetic ordering (designated phase 1) k=($\frac{1}{2}$,0,0) 
	\item a GdFeO$_3$ type structure (non charge ordered, orthorhombic cell,
	$a_{P}\sqrt{2}*a_{P}\sqrt{2}*2a_{P}$, space group \textit{Pbnm}) 
	associated to a ferromagnetic ordering k=(0,0,0)(designated phase 2).
\end{enumerate}
The model proposed leads to a good agreement with the data ($\chi^{2}=5.5$, phase1: relative quantity 92(1)$\%$,
 R$_{Bragg}$=2.39, R$_{magn}$=1.48; phase 2: relative quantity 92(1)$\%$, R$_{Bragg}$=1.78, R$_{magn}$=5.66) \cite{young}.
 The details of the nuclear and magnetic phases parameters are not given because this refinement is qualitative considering
 the complexity of the magnetic state of the \mbox{Bi$_{0.67}$Ca$_{0.33}$MnO$_{3}$}. Here, we mention just the main features that
are necessary in the present discussion. The precise magnetic structure refinement will be discussed in details elsewhere. \\
\indent The addition of the ferromagnetic ordering associated to the non charge ordered phase has been necessary to model
 properly the nuclear peaks. The major phase is the non charge ordered one, associated to the ferromagnetic ordering, and
 it represents about 90$\%$ of the diffracted volume. The correlation length extracted from the refinement for the ferromagnetic
 phase is about 10\AA . A correlation length of 620\AA\ has been extracted from the profile fitting with a Pseudo-Voigt function
 of the AFM Bragg peaks. The weak quantity of the charge ordered antiferromagnetic phase (about 10$\%$ of the compound) do not
 allow us to refine precisely the antiferromagnetic structure. However we can conclude that the antiferromagnetic structure
 is a pseudo CE type structure \cite{wollan}. The magnetic moments are along the c-axis with a week contribution
 in the ab-plan and they form ferromagnetic zig-zag chains ferromagnetically coupled along the c-axis and antiferromagnetically
 coupled in the ab-plan. \\The setting up of a "long range" antiferromagnetic (AFM) order below $T_N=150K$ 
is evidenced by the growth of AFM peaks corresponding to
 the propagation vector $k=(\frac{1}{2},0,0)$. The intensity of the AFM magnetic peaks saturates under
 about 40K (see the evolution of the integrated intensity of the ($\frac{1}{2}$,1,0) magnetic peak on Fig.10). 
\\In addition, a signal at small angles is observed in the NPD data (fig.9). 
Its magnitude increases when $T$ decreases (fig.11).
This signal is present over a "large" $Q$ range, from the lowest $Q$ up to $Q$ $=$ 0.5 $\AA^{-1}$. 
Its intensity is important even at 40K and it saturates below this temperature.
 Magnetic measurements have shown that spin glass properties appear under 39K. This temperature corresponds
 well to the one where we observe the saturation of the diffuse intensity. This small angle signal can therefore be interpreted as a ferromagnetic component with a very small correlation length.
In order to extract semi-quantitative results, we use the Guinier approximation, which is a Gaussian distribution of scattered
intensity. This corresponds to the case of dilute and non interacting clusters. This choice is dictated by the simplicity of the modeling, which gives a good fit in our restricted $Q$-range.
 In fact, the small Q-range reached with the G4.1 diffractometer, that is not an instrument dedicated to Small Angle Neutron Scattering measurement, do not justify a more sophisticated analysis. 
  The Guinier expression for $I_Q$ is a Gaussian of the form:
\begin{equation}
I_Q=I_0 exp-(R_G^3 Q^2/3)
\end{equation}
 
where $R_G$ is defined as the radius of gyration of the scattering object,
and $I_0$ is the intensity at $Q=0$.

With the hypothesis of an isotropic disorder, the scattering domain can be approximated by a sphere.
For a sphere of radius $R$, $R_G = \sqrt(3/5)  R$, giving here $R = 0.8 \pm 0.08 nm$.
This corresponds to about $1.5 \times$ the unit cell of the sample, and shows that the magnetic disorder is very local. 
Note that the correlation length $R_G$ does not depend significantly of the temperature (fig.11).
In contrast, $I_0$ increases when the temperature decreases, and makes a plateau below $T_{SG}$ (fig.11).  
$I_0$ depends of the magnetic contrast, the number and the volume of scattering objects.
Since $R_G$ is found to be temperature independent, the volume of the scattering objects can be taken as a constant.
 The increase of $I_0$ can be explained considering an increase of the number of scattering domains up to a 
dense regime which corresponds to the freezing temperature.

Note that to conciliate the magnetic properties and NPD results, it appears important to analyse both small angle 
scattering and Bragg scattering.
 Otherwise, some puzzling results, like the existence of spin glass properties without any trace of magnetic disorder (in the presence of only magnetic Bragg peaks), 
are reported \cite{goudurix}.

The existence of a Griffith phase in manganites
has recently received several experimental supports \cite{salamon}. A Griffith phase can be observed in the presence of a random
distribution of magnetic interactions. It can be expected in dilute ferromagnets if the temperature is between
the temperature of the pure (non dilute) system $T_p$
and the temperature where long range ordering emerges.
It was also argued that the competition between two ordered phases favors a Griffith phase \cite{burgy}.
Experimental signatures of a Griffith phase are the observation of magnetic clusters (with a long tail distribution)
 and an anomalous form of the susceptibility due to the non analytic thermodynamic functions, even far above $T_c$.
 In our case, we observe ferromagnetic entities (clusters) in a globally paramagnetic phase at a high temperature,
 and the emergence of magnetic order when the temperature is decreased.
Nevertheless, we do not observe a magnetic susceptibility exponent less than unity in the low field limit
 ($\chi^{-1} \propto (T-T_c)^{1-\lambda}$ with $0 \leq \lambda \leq 1$ ).
Moreover, the superparamagnetic like form temperature dependence of the low angle magnetic intensity and the constant correlation length 
that we observe are not in favor of a Griffith phase \cite{griffith}.

In conventional metallic spin glasses, the randomness arises from the dilution of magnetic ions in a paramagnetic matrix. 
An analogy can be proposed for our sample.
$BiMnO_3$ is known to be ferromagnetic when $Bi_{1-x}Ca_xMnO_3$ is antiferromagnetic up to $x>0.35$ \cite{maud}.
 The results presented in this paper show that 
the $Bi_{0.67}Ca_{0.33}MnO_3$ compound presents AFM domains with a correlation length of $620 \AA$ and ferromagnetic domains at the scale of the unit cell. 
Such a low ferromagnetic correlation length can be understood with a schematic picture of unit cells close to ferromagnetic $BiMnO_3$ which are randomly distributed in an antiferromagnetic (Bi,Ca)MnO$_3$ matrix (fig.13).
In our sample, the neutron diffraction data and the HREM pictures have indeed shown that
the occupation of the Bi and Ca sites is random. In the $(Ln,Ca)MnO_3$ with $Ln=Pr,Sm$ \cite{BR}, the stability limit of the charge ordering is linked
 to a phase separation between mesoscopic FM and AFM domains. In such a case, the application of magnetic fields induces a percolation of the FM domains \cite{damien}.
 In our sample, the ground state is a spin glass, which is quite different from a phase separation state.
 The ferromagnetism develops only at one unit cell, and the application of
 very low magnetic fields favors the AFM domains against the spin glass state. This can be relied on the peculiar robustness of the AFM state
 in $(Bi,Ca)MnO_3$ which shows a very high critical field ($B_c\approx 56 T$) \cite{kirste}. 
The percolation is considered to be essential to observe the CMR effect. In our sample, since the spin glass ground state is inconsistent with an evolution towards a percolation,
 the CMR effect is not expected. This is in agreement with our measurements (not shown), that do not report any CMR in $Bi_{0.67}Ca_{0.33}MnO_3$ up to 7T.

To conclude, $Bi_{0.67}Ca_{0.33}MnO_{3}$ shows a complex magnetic behavior at low temperature.
$Ac$ macroscopic measurements prove spin glass properties.
This implies the existence of magnetic frustration likely caused
 by the competition of ferromagnetic/antiferromagnetic exchange interactions.
 In addition, the susceptibility has symmetry breaking by field inversion in the spin glass state,
showing that magnetic order coexists with the spin glass order. 
Consistently, neutron scattering evidences a ferromagnetic correlation length of less than $1 nm$, which exists even in the presence of antiferromagnetic order.
 Contrarily to the phase separation observed at largest scale in other manganites, the spin glass ground state does not allow the CMR effect in $Bi_{0.67}Ca_{0.33}MnO_{3}$.
 
Acknowledgements:
We acknowledge Ch.Simon (CRISMAT) for useful comments on this work.

\newpage
\vskip 2 cm

\begin{figure}[htbp]
\includegraphics*[width=9cm]{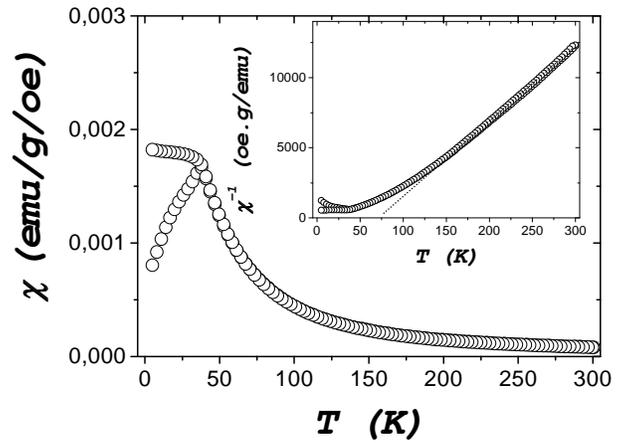}
\caption{Susceptibility of $Bi_{1-x}Ca_{x}MnO_3$ ($x=0.33$) as function of the temperature (Field cooling and zero field cooling with H = 100 Oe).
In the inset is shown the inverse of the susceptibility as function of the temperature.}
\end{figure}

\newpage

\vskip 2 cm

\begin{figure}[htbp]
\centering  
\includegraphics*[width=9cm]{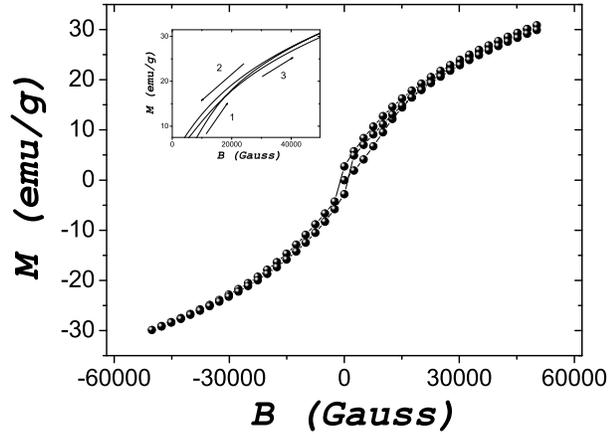}
\vskip 2 cm
\caption{Hysteresis cycle of the magnetization at $T\approx 2 K$. In the inset is shown the crossing of the up and down branches, what is characteristic
of slow relaxation during the measurements. }
\end{figure}

\vskip 2 cm
\begin{figure}[htbp]
\centering  
\includegraphics*[width=9cm]{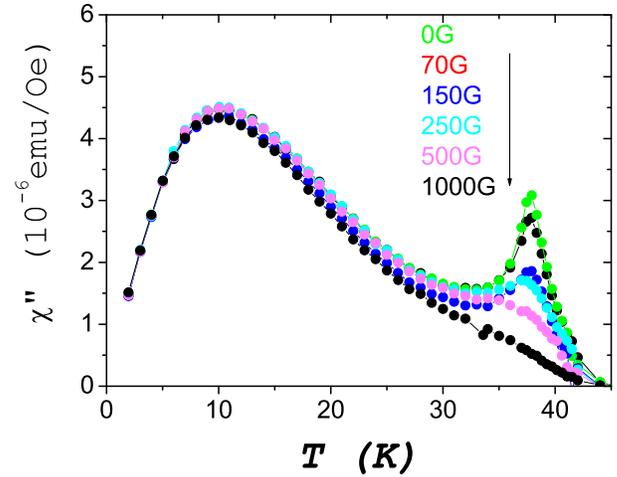}
\vskip 2 cm
\caption{Imaginary part of the susceptibility as a function of the temperature. Note the collapse of the dissipation peak at $T\approx 39 K$ for low to moderate applied field. Note also that for $T <30 K$, a second and broad peak emerges and that this latter is robust to the application of a magnetic field.}
\end{figure}

\vskip 2 cm

\begin{figure}[htbp]
\centering  
\includegraphics*[width=9cm]{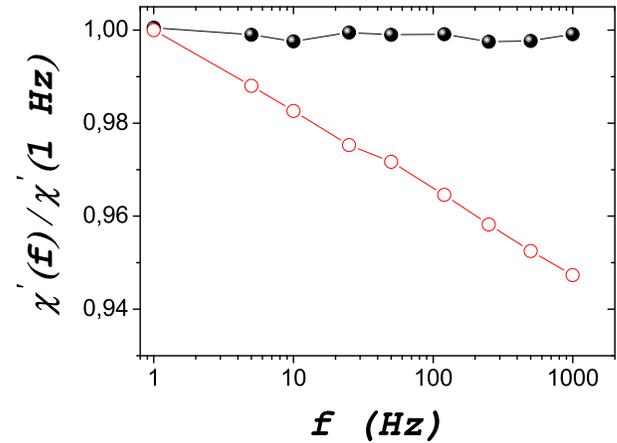}
\vskip 2 cm
\caption{Frequency dependence of the normalized real part of the ac susceptibility ($B_{dc}=10 G$). The top curve is in the paramagnetic regime regime ($T=45 K$),
 the bottom curve is at low temperature ($T=10 K$) and exhibits logarithm dependence typical of glassy states.}
\end{figure}

\vskip 2 cm

\begin{figure}[htbp]
\centering \includegraphics*[width=9cm]{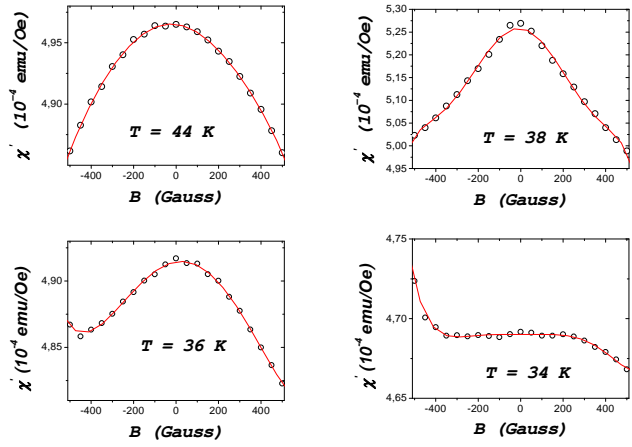}
\vskip 2 cm
 \caption{Low
field variation of $\chi^{'}$ for different temperatures. The solid line is a fit using the power-law expansion
(equation (2)). 
Note the symmetric magnetic field dependence in
the paramagnetic state close to the spin-glass transition ($T = 44
K$, $H^2$ correction) and the breaking of the symmetric
response implying the existence of a spontaneous
magnetic moment ($T= 36 K$).}
\end{figure}

\vskip 2 cm

\begin{figure}[htbp]
\centering \includegraphics*[width=9cm]{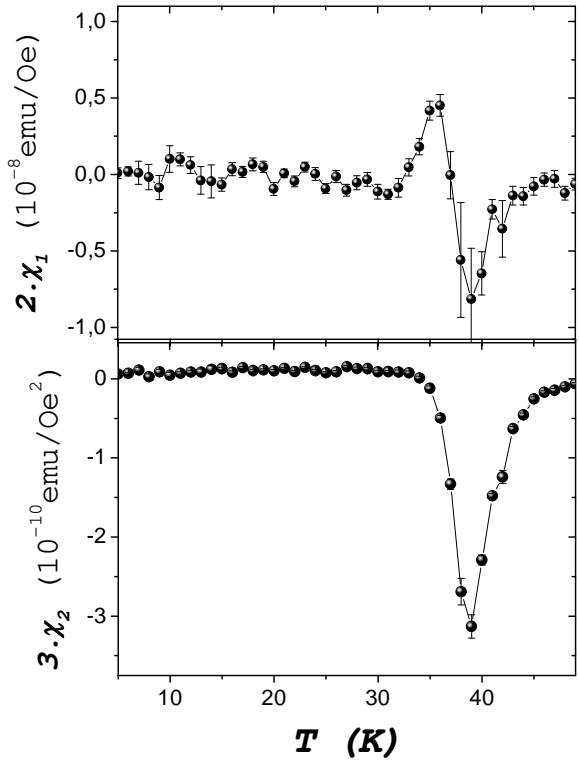}
\vskip 2 cm
\caption{Temperature variation of $\chi_1$ and $\chi_2$, the first parameters of the
non-linear susceptibility.}
\end{figure}

\vskip 2 cm

\begin{figure}[htbp]
\centering \includegraphics*[width=9cm]{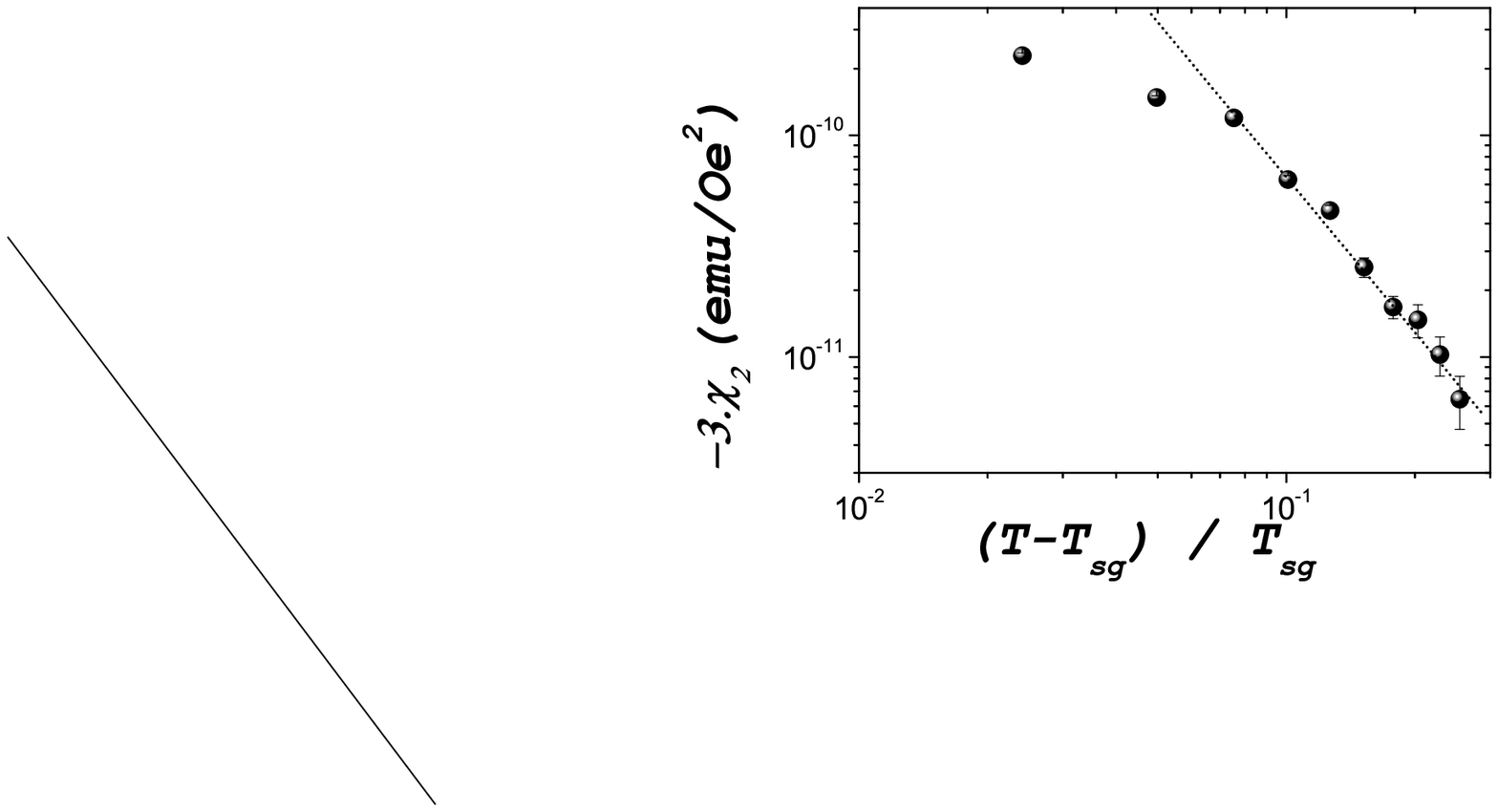}
\vskip 2 cm
\caption{ $|\chi_2|$ as a function of $\varepsilon = (T-T_{SG})/T_{SG}$ in a log-log scale, 
and showing the power-law dependence $|\chi_2| = \varepsilon^{\gamma}$ for $\varepsilon \geq 0.08$, with $\gamma = 2.4 \pm 0.1$.}
\end{figure}

\vskip 2 cm

\begin{figure}[htbp]
\centering \includegraphics*[width=9cm]{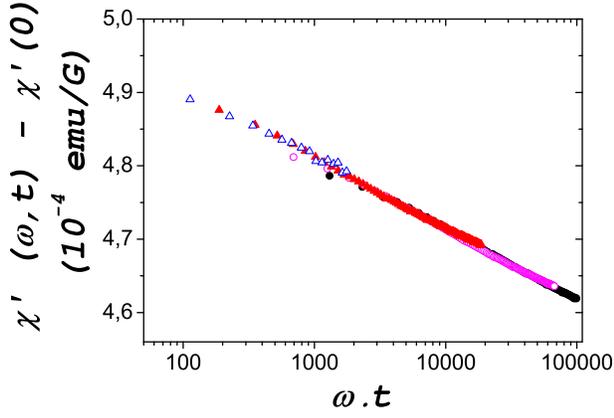}
\vskip 2 cm
\caption{Scaling of the linear part of the ac susceptibility, showing aging properties.
 The frequencies are $f=\omega  / 2\pi = 10, 5, 1$ and $0.07 Hz$, $B=10 G$.}
\end{figure}

\vskip 2 cm





\begin{figure}
\centering \includegraphics*[width=9cm]{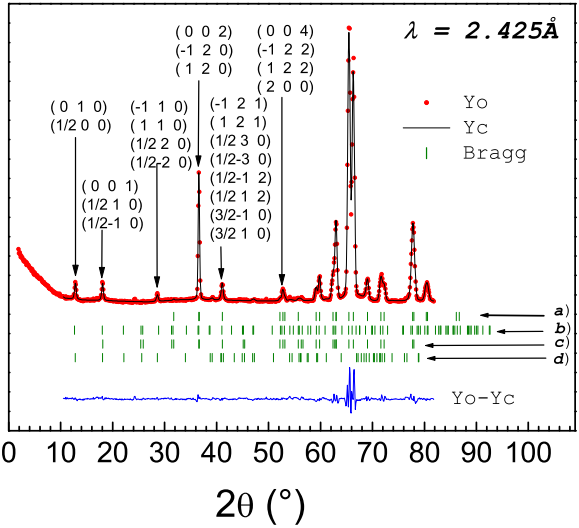}
\vskip 2 cm
\caption{\label{fig:fig1} \mbox{Bi$_{0.67}$Ca$_{0.33}$MnO$_{3}$} : Magnetic 
structure refined with the FullProf Suite of programs from neutron diffraction data collected at 1.5K on the LLB G4.1 diffractometer.
 The Bragg positions of the peaks (unit cell $a_{P}\sqrt{2}*2a_{P}\sqrt{2}*2a_{P}$) for each phase are indicated by small vertical line:
 a) for the non charge ordered phase (phase 1), b) for the charge ordered phase (phase 2),
 c) for the ferromagnetic order associated to the non charge ordered phase (magnetism of phase 1, propagation vector K=0),
 d) for the antiferromagnetic order associated to the charge ordered phase (magnetism of phase 2, propagation vector K=0.5 a$^{*}$)}
\end{figure}

\begin{figure}[htbp]
\centering \includegraphics*[width=9cm]{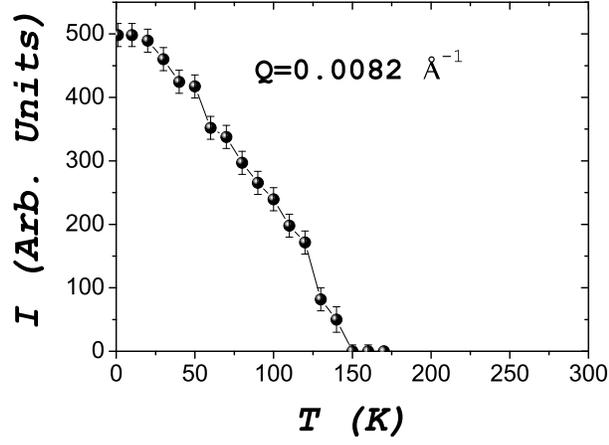}
\vskip 2 cm
\caption{Scattered intensity as a function of the temperature of the second AFM Bragg
 peak indexed ($\frac{1}{2}$,1,0) in the $a_{P}\sqrt{2}*2a_{P}\sqrt{2}*2a_{P}$ cell (with $a_{P}\simeq 3.9$ \AA the
 cubic perovskite cell parameter). The peak emerges at $T\approx 150K$,
 where a departure from the ideal Curie-Weiss law can be observed in the figure 1.}
 \end{figure}

\begin{figure}[htbp]
\centering \includegraphics*[width=9cm]{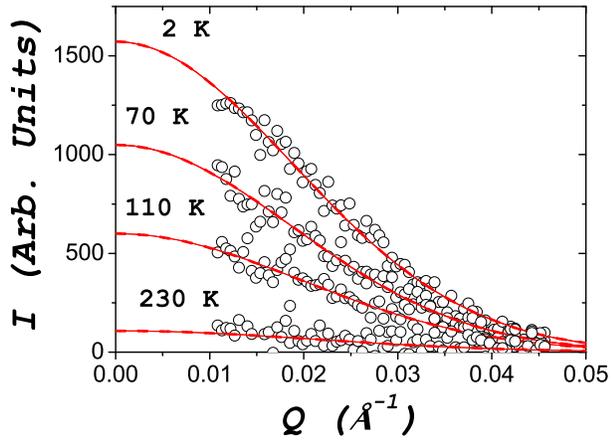}
\vskip 2 cm
\caption{Scattered intensity at small angles as function of $Q$ for different temperatures.
 The dashed line is a fit using the Guinier approximation (equation (5)).}
\end{figure}
\vskip 2 cm

\begin{figure}[htbp]
\centering \includegraphics*[width=9cm]{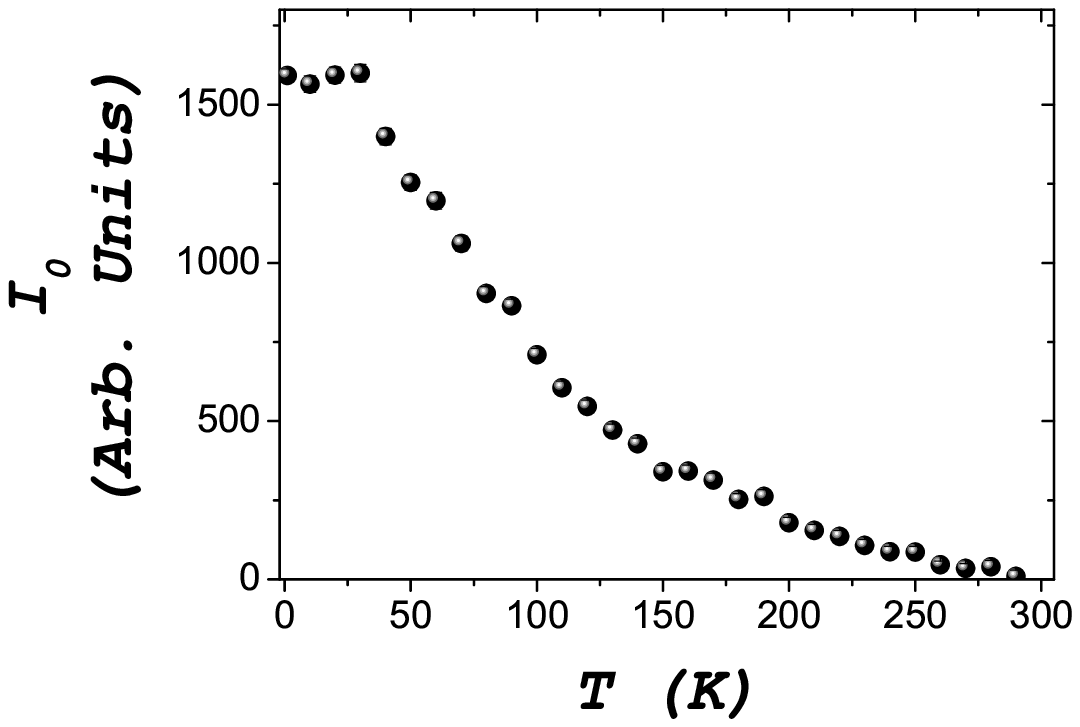}
\end{figure}

\begin{figure}[htbp]
\centering \includegraphics*[width=9cm]{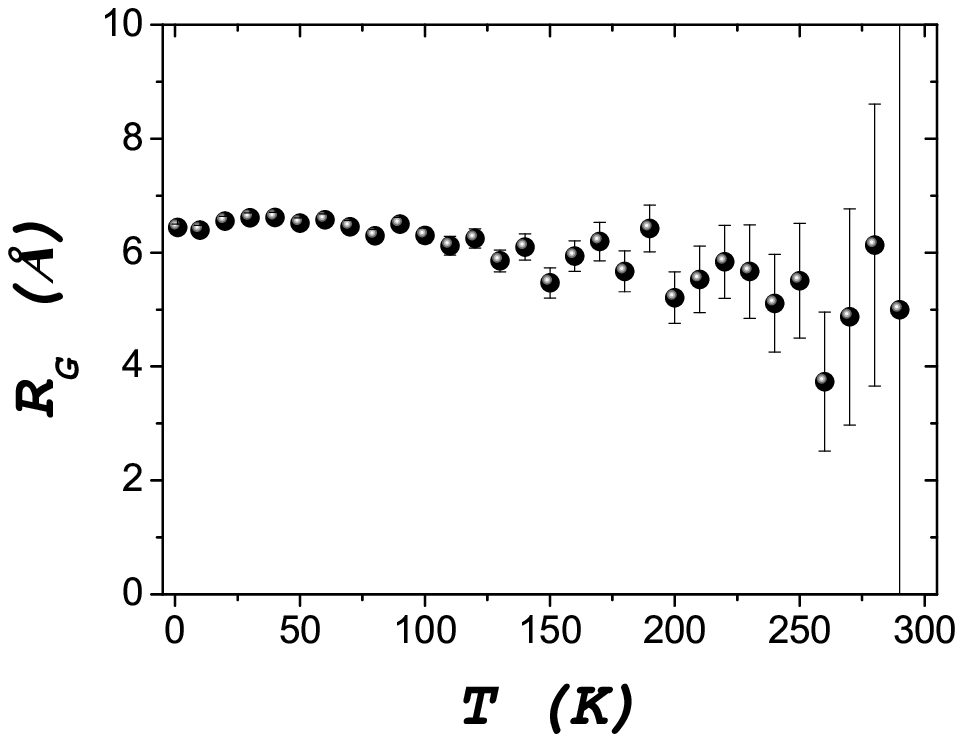}
\vskip 2 cm
\caption{a/Top: $I_0$ as function of the temperature and b/ Bottom: $R_G$,
 the gyration radius of the scattering domain, as function of the temperature.
 Note the plateau of $I_0$ below $T = 40 K$, corresponding to the spin glass domain.
 $R_G$ is roughly constant over the $1-300 K$ temperature range.}
\end{figure}

\vskip 2 cm

\begin{figure}[htbp]
\centering \includegraphics*[width=9cm]{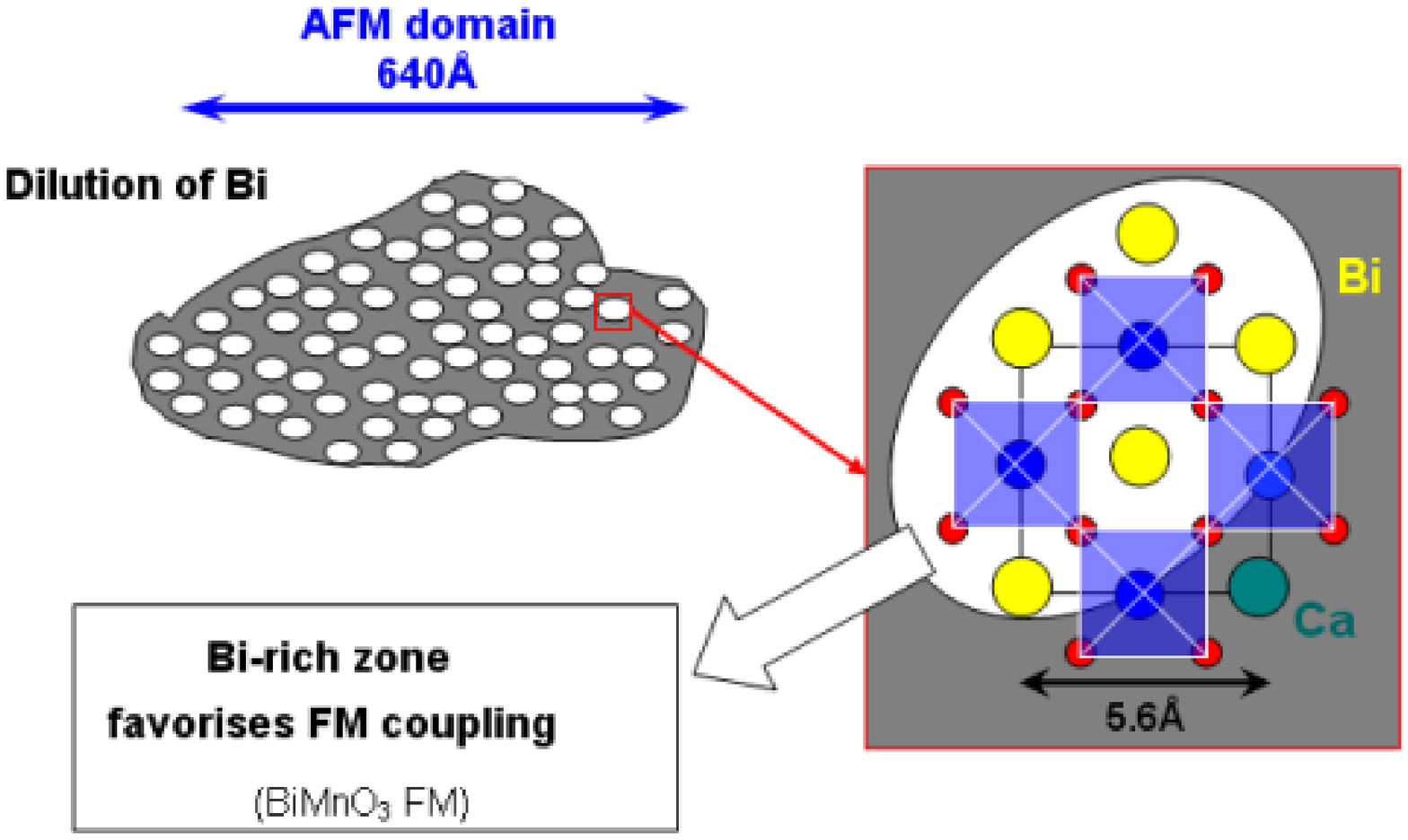}
\vskip 2 cm
\caption{Schematic picture of Bi rich unit cells, with dominant ferromagnetic interactions, randomly distributed in an antiferromagnetic (Bi,Ca)MnO$_3$ matrix.}
\end{figure}

\vskip 2 cm


\begin{references}
\label{sec:TeXbooks}
\bibitem{dagotto} E. Dagotto, T. Hotta, and A. Moreo, Physics Reports 344, 1 (2001).
\bibitem{order} P. Azaria, H.T. Diep, and H. Giacomini, Phys. Rev. Lett. 59, 1629 (1987).
\bibitem{bokov} V.A. Bokov, N.A. Grigoryan, and M.F. Bryzhina,
Phys. Stat. Sol. 20, 745 (1967).
\bibitem{troy} I. O. Troyanchuk, O. S. Mantytskaya, and A. N. Chobot, Physics of the Solid State 44, 2266 (2002).
\bibitem{degennes} P.-G. de Gennes, Phys. Rev. 118, 141 (1960).
\bibitem{binder} K. Binder and A. P. Young, Rev. Mod. Phys. 58, 801 (1986).
\bibitem{nail} S. Nair and A. Banerjee, Phys. Rev. B 68, 094408 (2003).
\bibitem{damien} D. Saurel, A. Brulet, A. Heinemann, C. Martin, S. Mercone, and C. Simon, Phys. Rev. B 73, 094438 (2006).
\bibitem{maud} M. Giot, A. Pautrat, O. Perez, C. Simon, M. Nevriva and M. Hervieu, Solid State Sciences 8, 1414 (2006).
\bibitem{mydosh} A. F. J. Morgownik, J. A. Mydosh, and C. L. Foiles
Phys. Rev. B 29, 4144 (1984).
\bibitem{saza} M. Sasaki, P. E. J$\ddot{o}$nsson, H. Takayama, and H. Mamiya, Phys. Rev. B 71, 104405
(2005).
\bibitem{natterman} T. Nattermann, Y. Shapir, and I. Vilfan, Phys. Rev. B 42, 8577
(1990).
\bibitem{pytte} E. Pytte and Y. Imry, Phys. Rev. B 35, 1465 (1987).
\bibitem{suzuki} M. Suzuki, Prog. Theor. Phys. 5 , 1151 (1977).
\bibitem{peak} S. Chikazawa, T. Bitoh, K. Ohba, M. Takamatsu, T. Shirane, Journal of Magnetism and Magnetic Materials 154, 59 (1996).
\bibitem{levy} L. L$\acute{e}$vy, Phys. Rev. B 38, 4963 (1988).
\bibitem{reent} H. P. Kunkel and G. Williams, J. Magn. Magn. Mater. 75, 98 (1988).
\bibitem{vincent} E. Vincent, J. Hammann, M. Ocio, J.P. Bouchaud, and L.F. Cugliandolo, in Complex Behaviour of Glassy Systems, Springer Lecture Notes in Physics Vol. 492,
edited by M. Rubi, (preprint cond-mat/9607224), pp. 184–219 and references therein.
\bibitem{reju} P. E. J$\ddot{o}$nsson, H. Yoshino, H. Mamiya, and H. Takayama, Phys. Rev. B 71, 104404 (2005).
\bibitem{ladieu} Frustration is necessary for spin glass properties, but the exact role of disorder is still in debate. See F. Ladieu, F. Bert, V. Dupuis, E. Vincent
and J. Hammann, J. Phys.: Condens. Matter 16, 735 (2004).
\bibitem{quantum} Mihai A. G$\hat{i}$rtu, Charles M. Wynn, Wataru Fujita, Kunio Awaga,
and Arthur J. Epstein, Phys. Rev. B 61, 4117 (2000). X. Obradors, J. Bassas, J. Rodriguez,
 J. Pannetier, A. Labarta, J. Tejada and F. J. Berry, J. Phys. Cond. Matter. 2, 6801 (1990). 
 \bibitem{wong} P. Wong, S. von Molnar, T. T. M. Palstra, J. A. Mydosh, H. Yoshizawa, S. M. Shapiro, and A.
Ito, Phys. Rev. Lett. 55, 2043 (1985). 
\bibitem{young} R. A. Young, The Rietveld Method, Oxford University Press (1995). 
\bibitem{wollan} E.O. Wollan and W.C. Koehler, Phys. Rev. 100, 5451 (1955).
\bibitem{goudurix} A. Maignan, U. V. Varadaraju, F. Millange and B. Raveau, Journal of Magnetism and Magnetic Materials 168, 237 (1997). 
R Rodriguez, A Fernandez, A Isalgue, J Rodriguez, A Labarta, J Tejada and X Obradors, J. Phys. C: Solid State Phys. 18, 401 (1985).
\bibitem{salamon} M. B. Salamon, P. Lin and S. H. Chun, Phys. Rev. Lett. 88, 197203 (2002).
\bibitem{burgy} J. Burgy, M. Mayr, V. Martin-Mayor, A. Moreo, and E. Dagotto
Phys. Rev. Lett. 87, 277202 (2001).
J. Burgy, E. Dagotto, and M. Mayr, Phys. Rev. B 67, 014410 (2003).
\bibitem{griffith} C. Magen, P. A. Algarabel, L. Morellon, J. P. Araújo, C. Ritter, M. R. Ibarra, A. M. Pereira, and J. B. Sousa, Phys. Rev. Lett. 96, 167201 (2006).
\bibitem{BR} B. Raveau and M. Hervieu, "Orbital and charge ordering in manganites" in New trends in the caracterisation of CMR-manganites and related materials,
edited by K. B\"{a}rner, 79-137 (2005).
\bibitem{kirste} A. Kirste, M. Goiran, M. Respaud, J. Vanaken, J. M. Broto, H. Rakoto, M. von Ortenberg, C. Frontera, and J. L. Garc\'{i}a-Mu\~{n}oz, Phys. Rev. B 67, 134413 (2003).
\end{references}
\end{document}